\documentclass[aps,preprint,showkeys,superscriptaddress]{revtex4}
\usepackage{times}
\usepackage{amsmath}
\usepackage{epsfig}
\usepackage{dcolumn}
\usepackage{bm}
\usepackage[utf8]{inputenc}
\usepackage{graphicx}
\usepackage{float} 
\usepackage{subfigure}

\usepackage{CJKutf8}

\usepackage{color}

\newcommand{\pp}{$p$}
\newcommand{\pim}{$\pi^-$}
\newcommand{\nbar}{$\bar{n}$}
\newcommand{\jpsi}{$J/\psi$}

\newcommand{\pres}{$\sigma_p$}
\newcommand{\tres}{$\sigma_\theta$}
\newcommand{\phires}{$\sigma_\phi$}

\begin{document}

\title{\boldmath An approach to study interactions of antineutrons with CsI \\
at a \jpsi~factory}
\author{Si-Cheng Yuan}\email{yuansicheng@ihep.ac.cn} 
\author{Liang-Liang Wang}\email{llwang@ihep.ac.cn} 
\author{Wei-Dong Li}\email{liwd@ihep.ac.cn} 

\affiliation{Institute of High Energy Physics, Chinese Academy of Sciences, Beijing 100049, China}
\affiliation{University of Chinese Academy of Sciences, Beijing 100049, China}

\date{\today}

\begin{abstract}

Cesium Iodide (CsI) crystals are widely used in high-energy physics 
for their scintillation properties that enable detection of charged and neutral 
particles via direct and indirect ionization and form the basis of electromagnetic
calorimeters. However, knowledge of antineutron interactions with CsI 
is limited due to the difficulty of obtaining sources of antineutron
of sufficient intensity and energy definition. 
As antineutron are abundantly produced by many processes it would be 
particularly useful to improve understanding of the interactions of antineutrons
with CsI crystals. 

We propose to use the decay $J/\psi\to p\pi^-\bar{n}$ at the BEPCII $J/\psi$ factory
as a source of antineutrons using the BESIII detector with a CsI target added between 
the beam pipe and the detector. The BESIII Monte Carlo simulation with varying thicknesses of CsI target 
is used to validate the approach and optimize the target thickness. 
Selecting $p\pi^-$ charged particle tracks from the Monte Carlo we obtain clean antineutron samples 
with well defined momentum and direction. 
The selection efficiency, momentum and angular resolutions, as well as the interaction probability 
between antineutron and the CsI target are estimated. 

As the CsI thickness is increased more antineutron CsI interactions are obtained,
however the quality of the $p\pi^-$ selection is degraded. The Monte Carlo study yields
an optimum thickness that balances these effects. 
This approach can be applied to similar experiments with other types of target materials to measure baryons 
such as liquid hydrogen/deuterium and $\Lambda/\Xi$ hyperons.
%
%
\end{abstract}


\maketitle

\section{Introduction}

Many decays of $J/\psi$, $\psi(2S)$, $\Lambda_c$, and $B$~\cite{pdg} have final states that include the antineutron, but the 
detection of antineutron is limited as it is a neutral particle that does not directly cause ionisation. 
In some cases the antineutron is inferred from the recoil of the other final state particles~\cite{BESIII:2012imn,CLEO:2008aum}.
Alternatively it is identified by examining the shower produced by its interaction with a calorimeter~\cite{BESIII:2021tbq}.
Detection with a calor
Reliable Monte Carlo (MC) simulation of the antineutron and the calorimeter material interaction is required to obtain a correct selection efficiency and effective background suppression. 

Cesium Iodide (CsI) has high luminous intensity, high luminescence, small radiation length and stable chemical property. It is widely used as the interaction matter of the electromagnetic calorimeters for particle and nuclear physics experiments, such as BESIII~\cite{bes3}, CLEO-c~\cite{cleoc}, and Belle~II~\cite{belle2}. In these experiments, the interaction of antineutron and CsI is simulated with {\sc geant4}~\cite{geant4}, however the agreement between data and MC simulation is usually poor and data driven method is used when reliable estimation of the efficiency is needed but the precision is still limited and the full event shape information can hardly be used~\cite{Liu:2021rrx}.  

To achieve better precision for the antineutron involved measurements, more reliable MC simulation is needed, and knowledge on the antineutron and the calorimeter material interaction is necessary while so far it is not available due to extremely rare experimental data on antineutron interaction with any material. The main reason of the lack of data is due to the fact that antineutron beams with suitable intensity and momentum control are typically difficult to be obtained. The best antineutron sources obtained so far are from BNL E-767~\cite{E767} and the CERN OBELIX~\cite{OBELIX} experiments with limited statistics and momentum range ($<500$~MeV/$c$), and the experimental measurements are more related to the interaction with light nuclei (for a review, see~\cite{nbarPhysics}).

Because of the high cross section of $e^+e^-\to J/\psi$ and large decay branching fractions of $J/\psi$ to antineutrons like $J/\psi\to p\pi^-\bar{n}$, super \jpsi~factories are proposed to provide antineutron sources by tagging the accompanying particles like $p$ and $\pi^-$ in $J/\psi\to p\pi^-\bar{n}$  and to perform rich experiments related to nuclear and particle physics by placing specific custom-made targets just outside of the beam pipe~\cite{hypronProjectileFromJpsi}. Besides comparable or higher statistics (depending on luminosity) with respect to the existing antineutron sources, the maximum momentum of antineutrons from $J/\psi\to p\pi^-\bar{n}$ can reach $1174$~MeV/$c$ providing an unique chance to perform studies beyond 500~MeV/$c$.

The BESIII experiment~\cite{bes3} is an $e^+e^-$ annihilation experiment which can produce large amount antineutrons to some measurements never done before. We propose a measurement of the antineutrons and CsI interactions with the antineutron produced in $e^+e^-\to J/\psi\to p\pi^-\bar{n}$ by putting CsI target just outside of the beam pipe of the BESIII experiment. This will allow a measurement of many final states of the interactions and the data will be essential for the experiments, whether in operation or being proposed, which use CsI as the electromagnetic calorimeter. 

In this article, we make a specific proposal to study the interaction between antineutrons and CsI, using antineutron source from $J/\psi$ decays and a CsI target. By taking BESIII~\cite{bes3} as a demonstration, we perform quantitative validations of the proposal based on full MC simulation. Adding a CsI target in the full simulation of BESIII is described in Sec.~\ref{sec:target}, the simulation results of single $p$, single $\pi^+$, single $\bar{n}$, and $J/\psi\to p\pi^-$\nbar\ events are shown in Sec.~\ref{sec:results}, and the conclusion and perspectives are discussed in Sec.~\ref{sec:conclusion}. 

\section{\boldmath The ${\rm CsI}$ target and antineutron source at BESIII}
\label{sec:target}
    
The BESIII~\cite{bes3} at BEPCII~\cite{bepc2} is an $e^+e^-$ collision experiment running in the tau-charm energy region. Rich physics topics are covered by the BESIII experiment, including light hadron spectroscopy, charmonium and charmoniumlike exotic states, charm physics, QCD studies with light meson decays, tau physics and so on~\cite{bes3physics,bes3physicsFuture}. Due to the large production rate of $e^+e^-\to J/\psi$, it is also a $J/\psi$ factory. The BESIII detector, as described in detail in Ref.~\cite{bes3}, consists of a multi-layer drift chamber (MDC) filled with helium-based gas outside of a beryllium beam pipe, time-of-flight (TOF) counters made of plastic scintillators, an electromagnetic calorimeter made of CsI(Tl) crystals, a superconducting magnet providing a field of 1~Tesla, and a muon system made of resistive plate chambers.
    
The outer radius and the length of the BESIII beam pipe are 33.7~mm and 296~mm, respectively. The inner radius of MDC is 59.2~mm. The gap between the beam pipe and MDC is filled with air. Adding some matter as the target in this gap does not affect the beam or the detector. The proposed target material, CsI (we use pure CsI and neglect the Thallium in the simulation), has one Cesium atom and one Iodide atom inside, and a density of $4.51$~g/cm$^3$. The CsI target is formed into a tube shape with the same length as the outer beryllium tube of the beam pipe and attached to it. The thickness of additional CsI is the most important parameter that affects the antineutron production. On the one hand, a thicker CsI can cause more antineutrons to interact with it, increases the number of samples, and improves precision of the measurement; on the other hand, thicker CsI reduces the efficiency and resolution of the tagging particles (proton and $\pi^-$), thus reduces the efficiency, the momentum and angular resolutions of the tagged antineutrons. For these reasons, the determination of the thickness is a trade off the quantity and quality, and is studied in detail in the following sections.

The BESIII experiment~\cite{bes3} can collect 10~billion \jpsi\ events every year, of which $(2.12\pm 0.09)\times 10^{-3}$ are $p\pi^-\bar{n}$~\cite{pdg}. By identifying the proton and the $\pi^-$ originating from the interaction point and checking the recoiling mass of the $p\pi^-$ system, one can see clear antineutron signal at a recoil mass of 0.94~GeV/$c^2$ over very low background and obtain a sample of antineutrons with more than 99\% purity. The selection of events with antineutron interaction with additional CsI tube outside of the beam pipe are basically the same, with an additional requirement that the charged tracks excluding \pp\ and \pim\ originate from a secondary vertex located in the added CsI tube. 

The goal of our designed experiment is to study the antineutron and CsI interactions, and the quantity and quality of the tagged antineutrons and the quantity of the antineutron and CsI interactions are essential factors to be investigated. As mentioned in the previous section, quantity refers to how many antineutrons can be produced and the efficiency of the tagging process, while quality refers to the resolutions, including momentum resolution and angular resolution of the predicted antineutrons, both of which are important for the following physics analysis. 
    
Throughout the study, the official BESIII software {\sc boss} is used for simulation and reconstruction~\cite{boss,bes3physics,bes3hough}. An additional CsI tube needs to be introduced to the BESIII detector. In the simulation, it needs to be declared when defining the detector structure, and in the reconstruction, it needs to be added before track fitting with the Kalman filter~\cite{kalFit} to take into account the energy loss and multi-scattering effects properly. In order to avoid unnecessary errors caused by inconsistency, a global service is implemented to assure that identical materials and geometries appear in simulation and reconstruction. Single proton, $\pi^-$, and antineutron particle samples, as well as $J/\psi\to p \pi^- \bar{n}$ samples are generated by specifying different parameters including the additional CsI tube thickness, momentum and direction of the single particles. We study the efficiencies and resolutions in all these cases and determine the optimized thickness of the additional CsI tube.

\section{Study of simulated samples}
\label{sec:results}

Firstly, we study how the additional material caused by the CsI layer will affect the performance of track reconstruction for different types of particles such as protons, pions, and antineutrons. And then, we will investigate the combined effect on the event selection of $J/\psi\to p \pi^- \bar{n}$ . For simplicity, background particles introduced by the accelerator beam haven’t been taken into account in the simulation.

\subsection{Study of single proton and pion samples}

The tagged particles in our study are \pp\ and \pim. Compared with \pim, \pp\ will lose more energy in the same material due to its large mass, thus at low momentum the efficiency of \pp\ is much lower than that of \pim. For this reason, different momentum ranges are set in \pp\ and \pim simulations. The minimum momentum is 0.3~GeV/$c$ for \pp\ and 0.1~GeV/$c$ for \pim; meanwhile, the maximum momentum is 1.2~GeV/$c$ for both \pp\ and \pim. As for the polar angle $\theta$, the range of $\cos\theta$ is set to $[0,~0.93]$, half of the detector coverage. The particles are distributed uniformly in both momentum and $\cos\theta$. In total, $10^6$ events of each \pp\ and \pim\ sample are simulated.
        
For tag particles, we are mainly concerned with the efficiency of reconstruction and the resolutions of momentum and direction. Kalman filter is applied to the reconstructed tracks and the efficiency is defined as the number of selected tracks divided by the number of generated ones. After selection, the tracks are divided into several momentum and polar angle bins with a width of 0.05~GeV/$c$ and 0.05~radian, respectively. The difference between the momentum and polar angle reconstructed from Kalman filter and those from MC truth for each track is calculated to measure the resolution of the reconstruction. Thereupon, a Gaussian fit is applied and finally the resolutions of momentum and polar angle measurements are obtained.
        
\begin{figure}[htbp]
        	\centering  
        	\subfigbottomskip=2pt 
        	\subfigcapskip=-5pt 
        	\subfigure[]{
        		\includegraphics[width=0.4\linewidth]{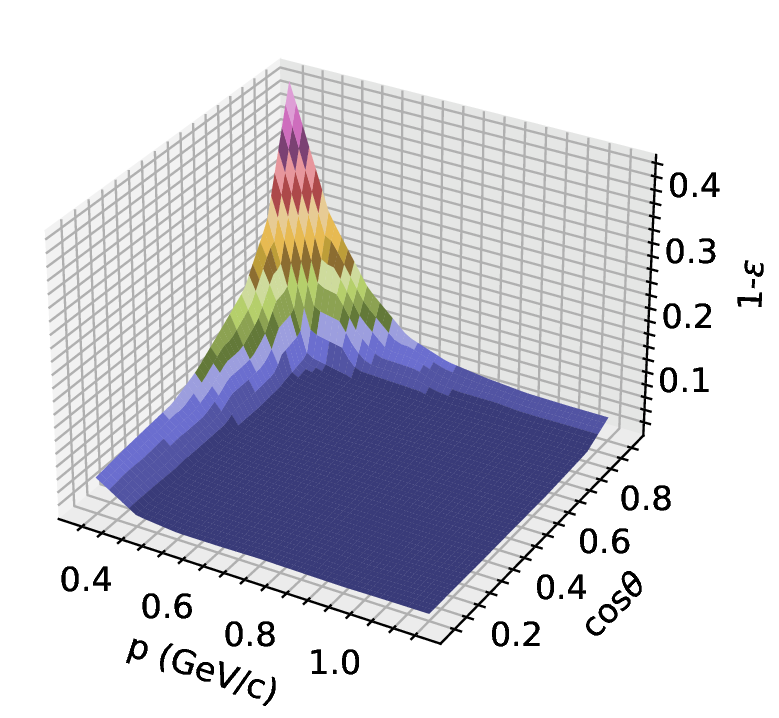}}
        	\subfigure[]{
        		\includegraphics[width=0.4\linewidth]{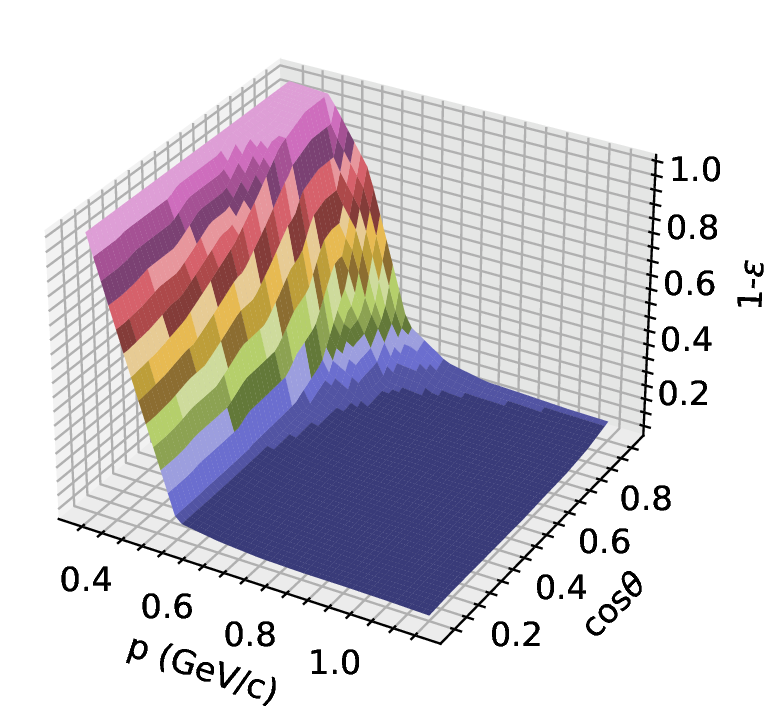}}
        	\subfigure[]{
        		\includegraphics[width=0.4\linewidth]{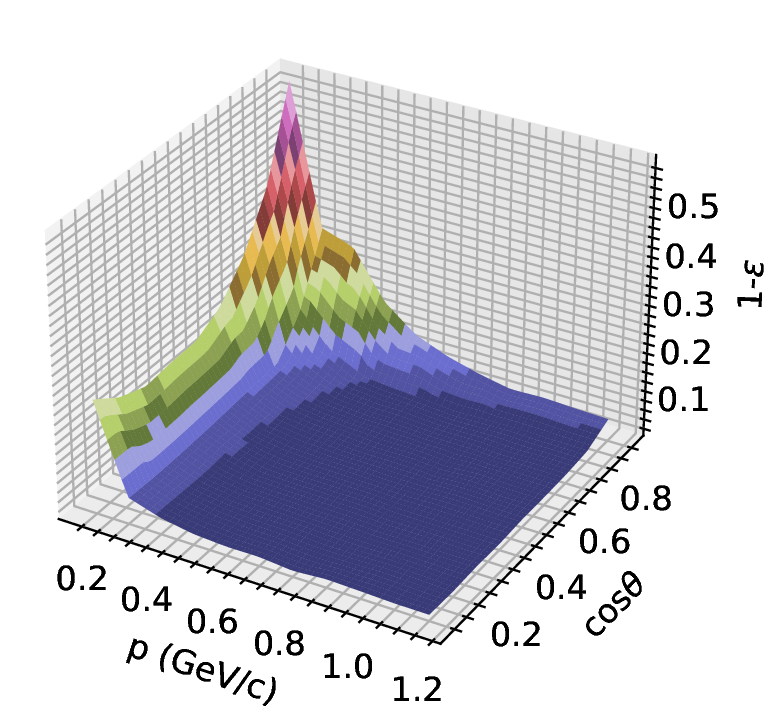}}
        	\subfigure[]{
        		\includegraphics[width=0.4\linewidth]{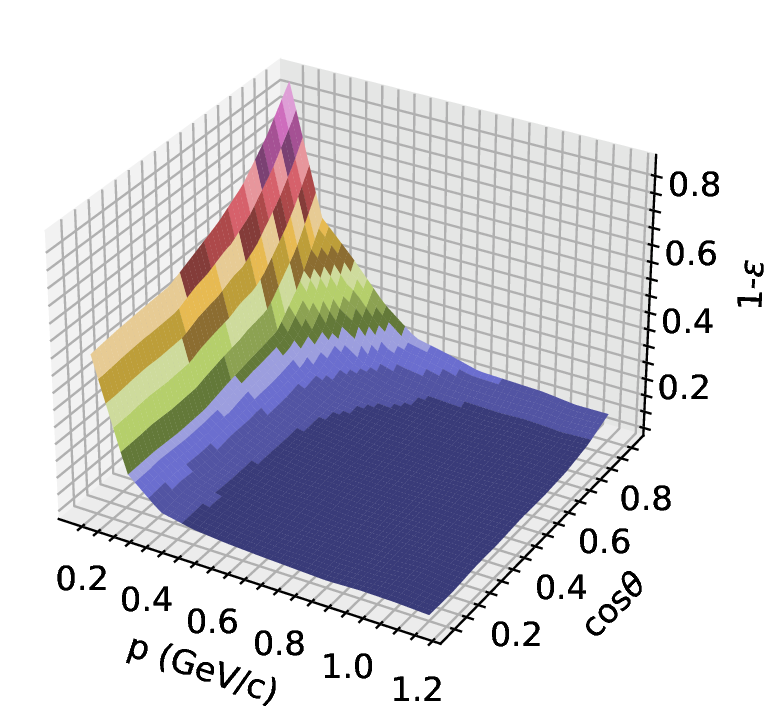}}
        	\caption{ Loss of tracking efficiency as a function of momentum and $\cos\theta$ (a) for protons without a CsI layer(b) for protons with  adding  a 10 mm thick CsI layer (c) for pions without a CsI layer  (d) for pions with adding  a 10 mm thick CsI layer.  }
        	\label{tracking}
        \end{figure}

Figure~\ref{tracking} shows the comparison of the proton and $\pi^-$ tracking efficiency loss ($1-\varepsilon$) without and with 10~mm CsI as a function of momentum and $\cos\theta$. As expected, the smaller the momentum is, the larger the angle is, the greater the reduction of the reconstruction efficiency with CsI added compared with that without CsI. The effect of added CsI tube on \pp\ is far greater than that of \pim\ due to the large $dE/dx$ for low momentum protons.
        
\begin{figure}[htbp]
        	\centering  
        	\subfigbottomskip=2pt 
        	\subfigcapskip=-5pt 
        	\subfigure[]{
        		\includegraphics[width=0.45\linewidth]{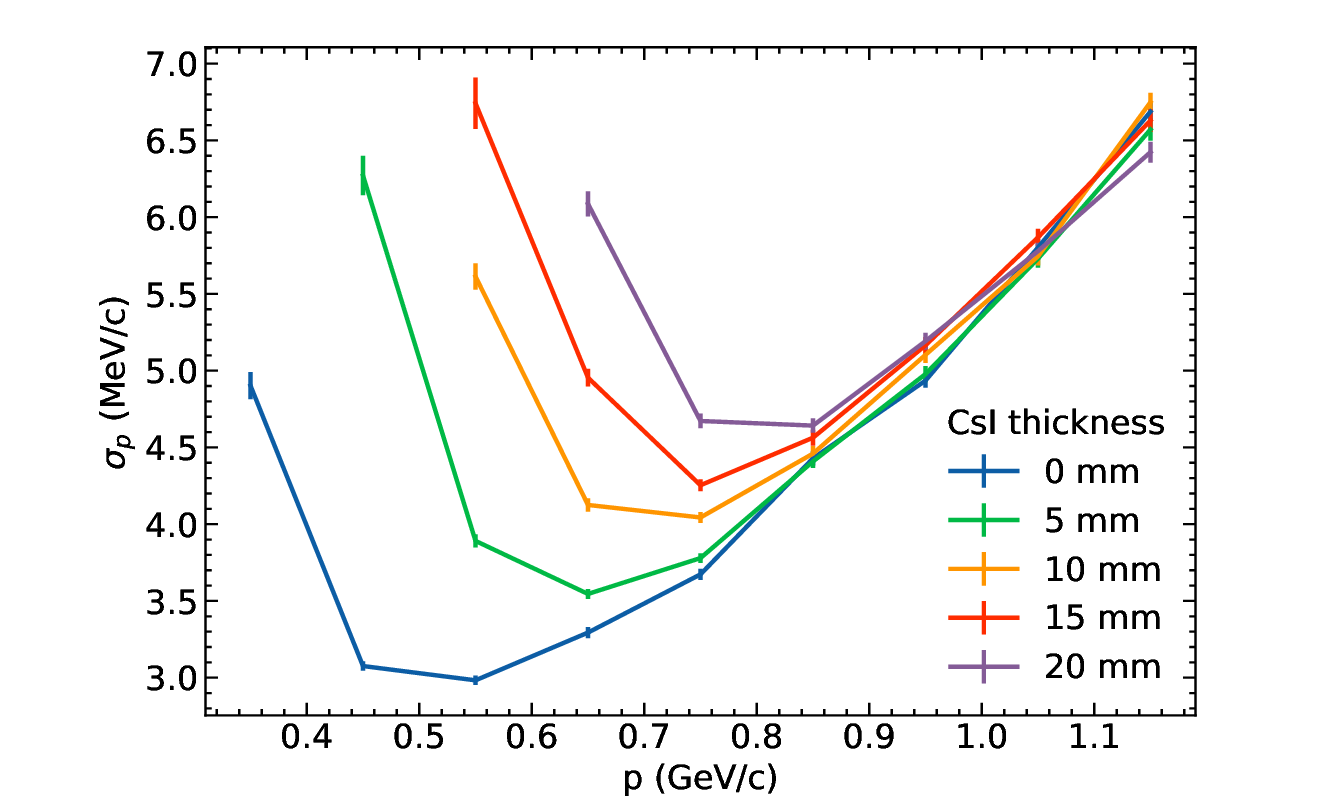}}
        	\subfigure[]{
        		\includegraphics[width=0.45\linewidth]{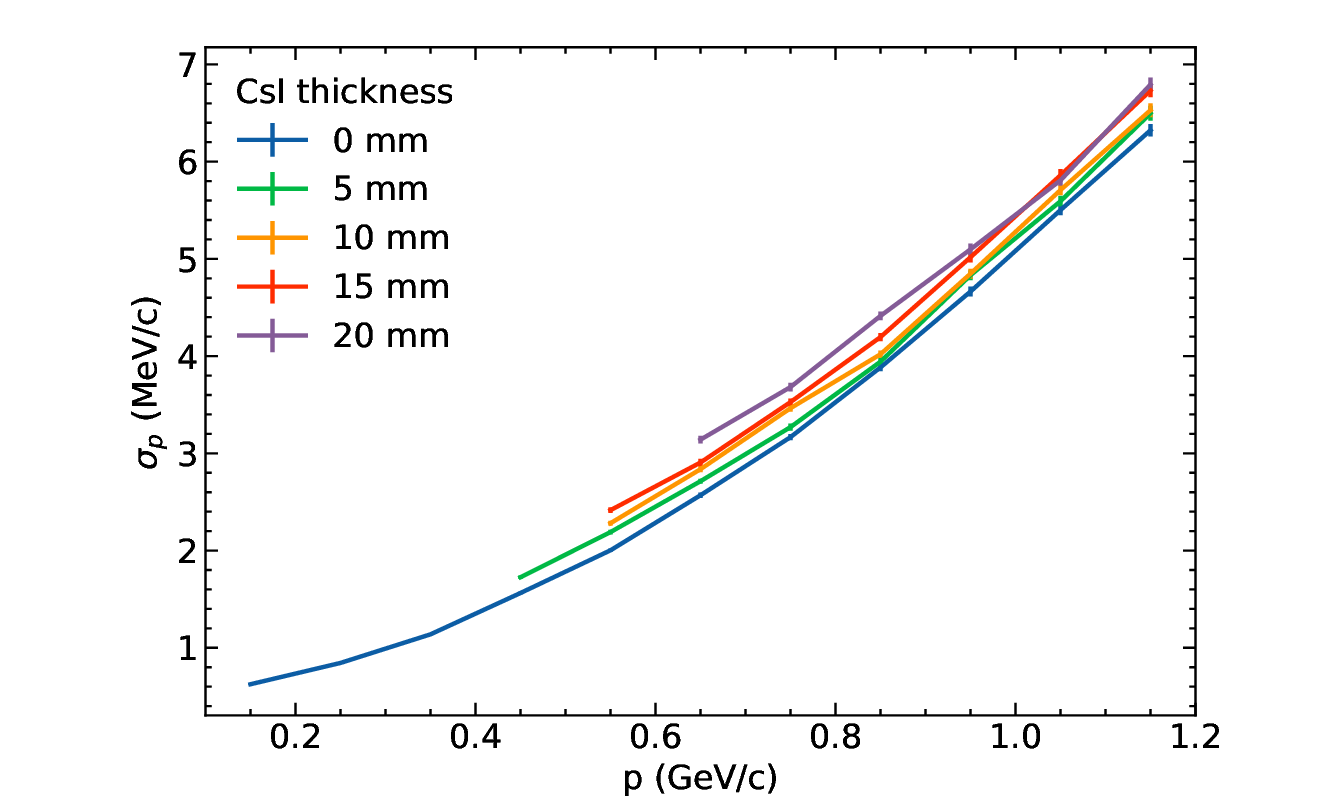}}
        	\caption{Momentum resolution \pres~ as a function of momentum p  for (a) protons (b) pions generated within the range of $0.1 < \cos\theta < 0.2$ for adding a CsI layer with the thickness of 0, 5, 10, 15, and 20 mm.}
        	\label{fig:p_resolution}
        \end{figure}

\begin{figure}[htbp]
        	\centering  
        	\subfigbottomskip=2pt 
        	\subfigcapskip=-5pt 
        	\subfigure[]{
        		\includegraphics[width=0.45\linewidth]{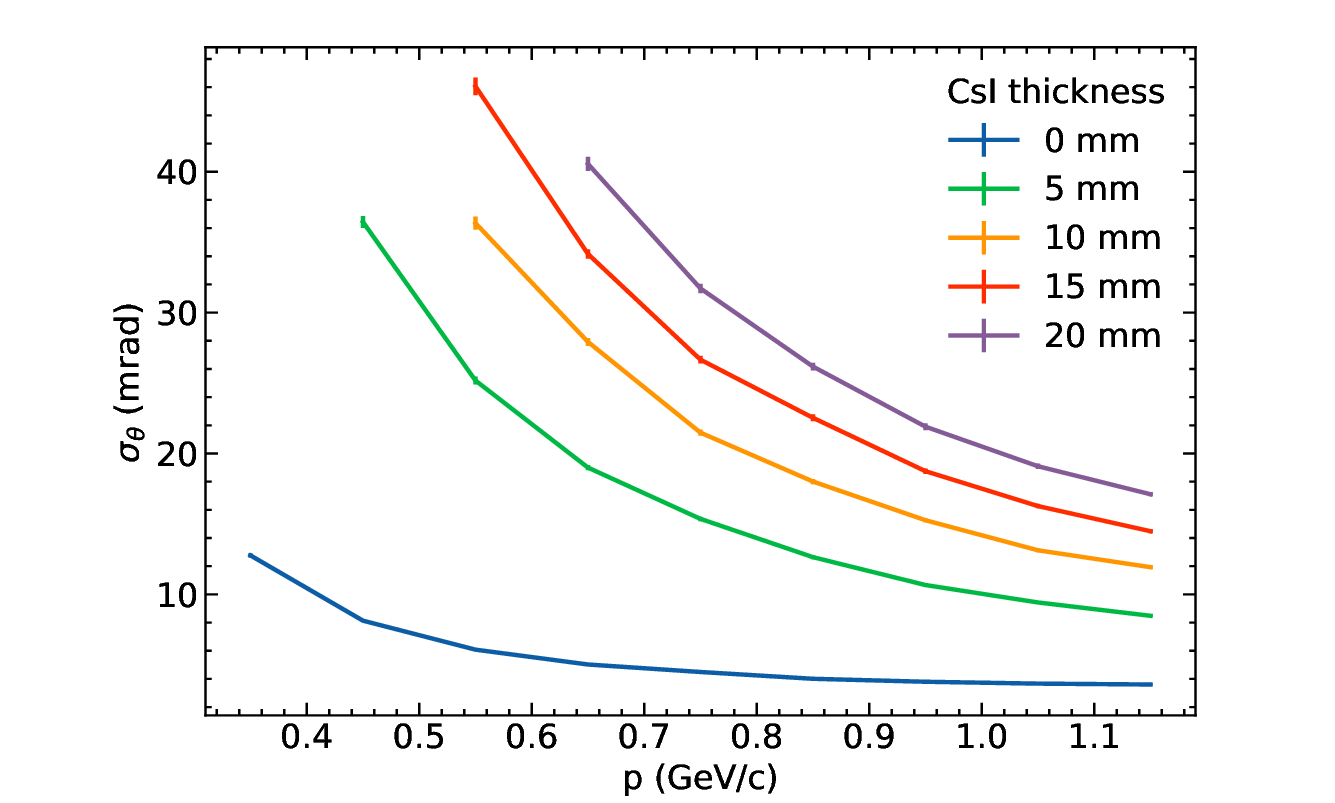}}
        	\subfigure[]{
        		\includegraphics[width=0.45\linewidth]{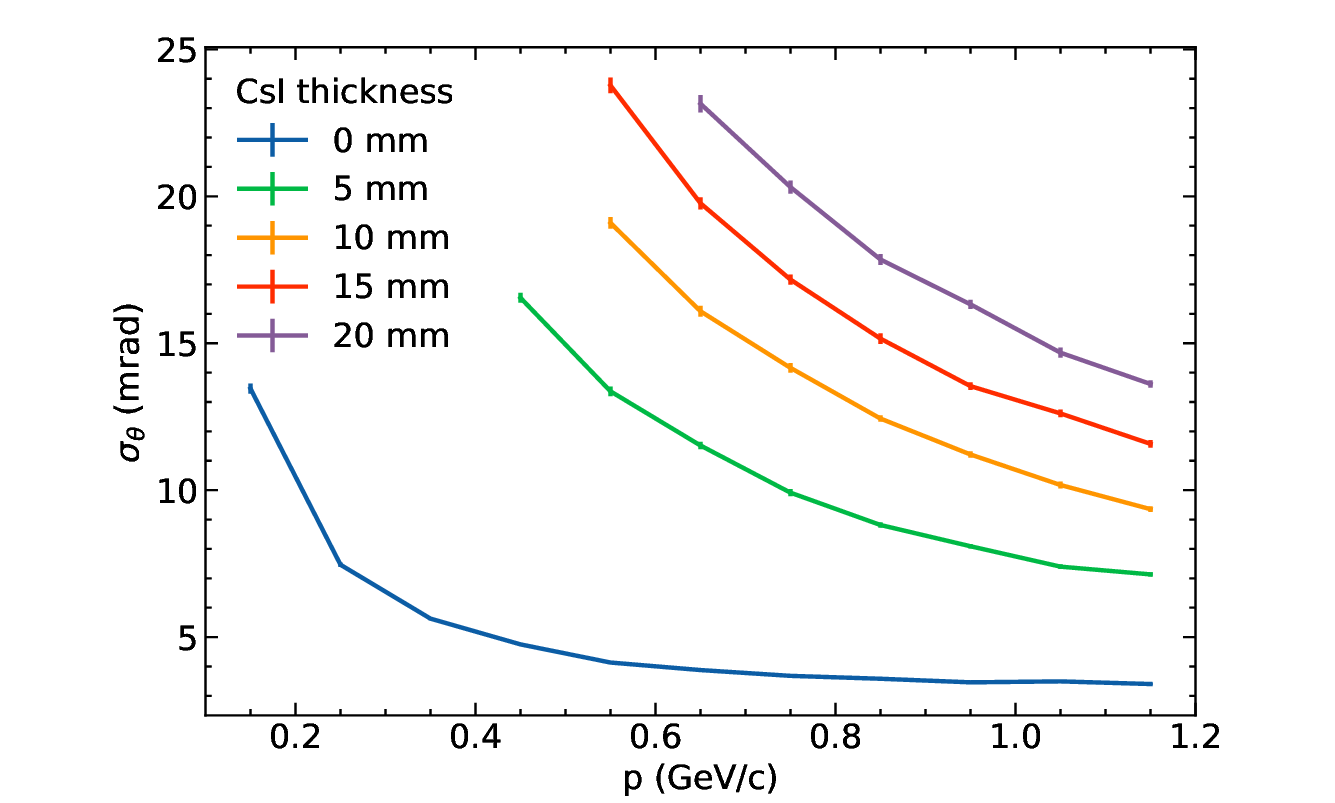}}
        	\caption{Polar angle resolution, \tres, as a function of of momentum p  for (a) protons (b) pions generated within the range of $0.1 < \cos\theta < 0.2$ for adding a CsI layer with the thickness of 0, 5, 10, 15, and 20 mm. }
        	\label{fig:t_resolution}
        \end{figure}

\begin{figure}[htbp]
        	\centering  
        	\subfigbottomskip=2pt 
        	\subfigcapskip=-5pt 
        	\subfigure[]{
        		\includegraphics[width=0.45\linewidth]{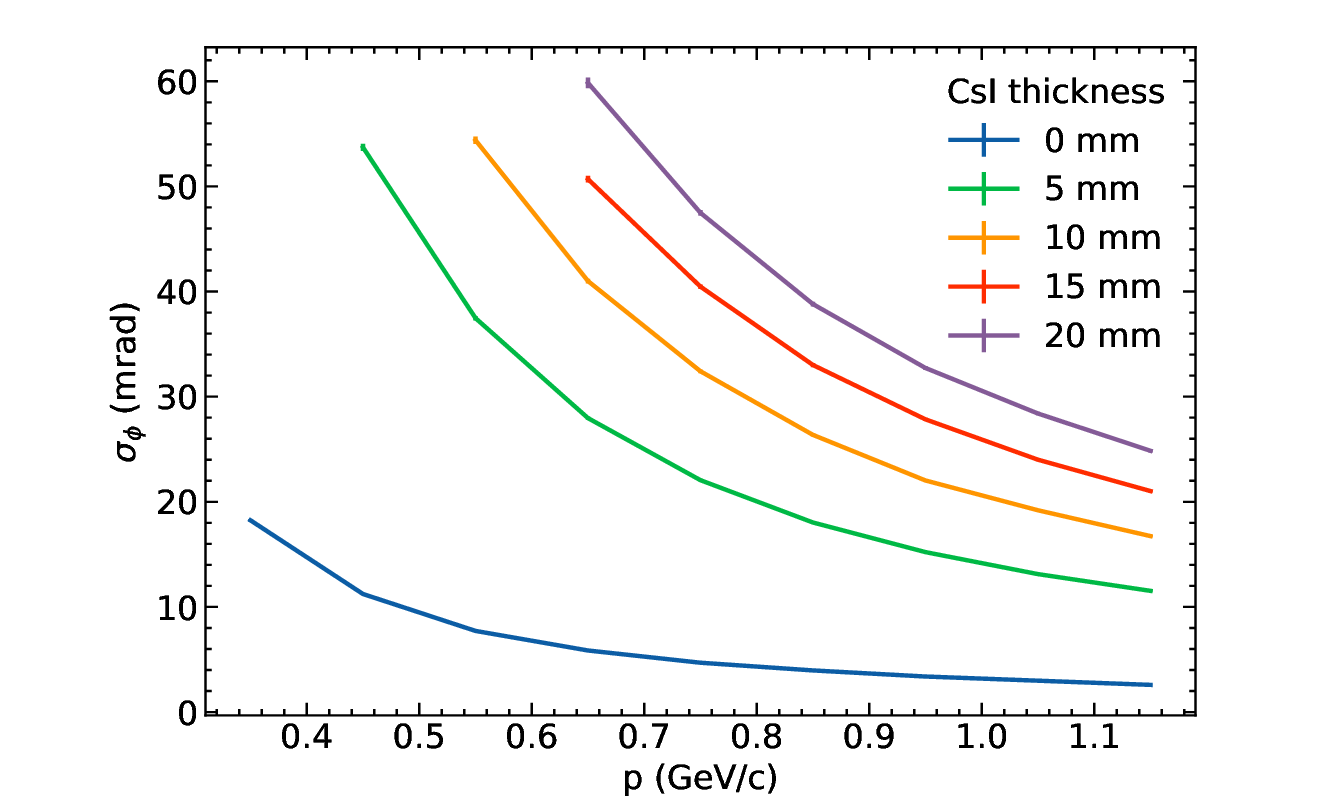}}
        	\subfigure[]{
        		\includegraphics[width=0.45\linewidth]{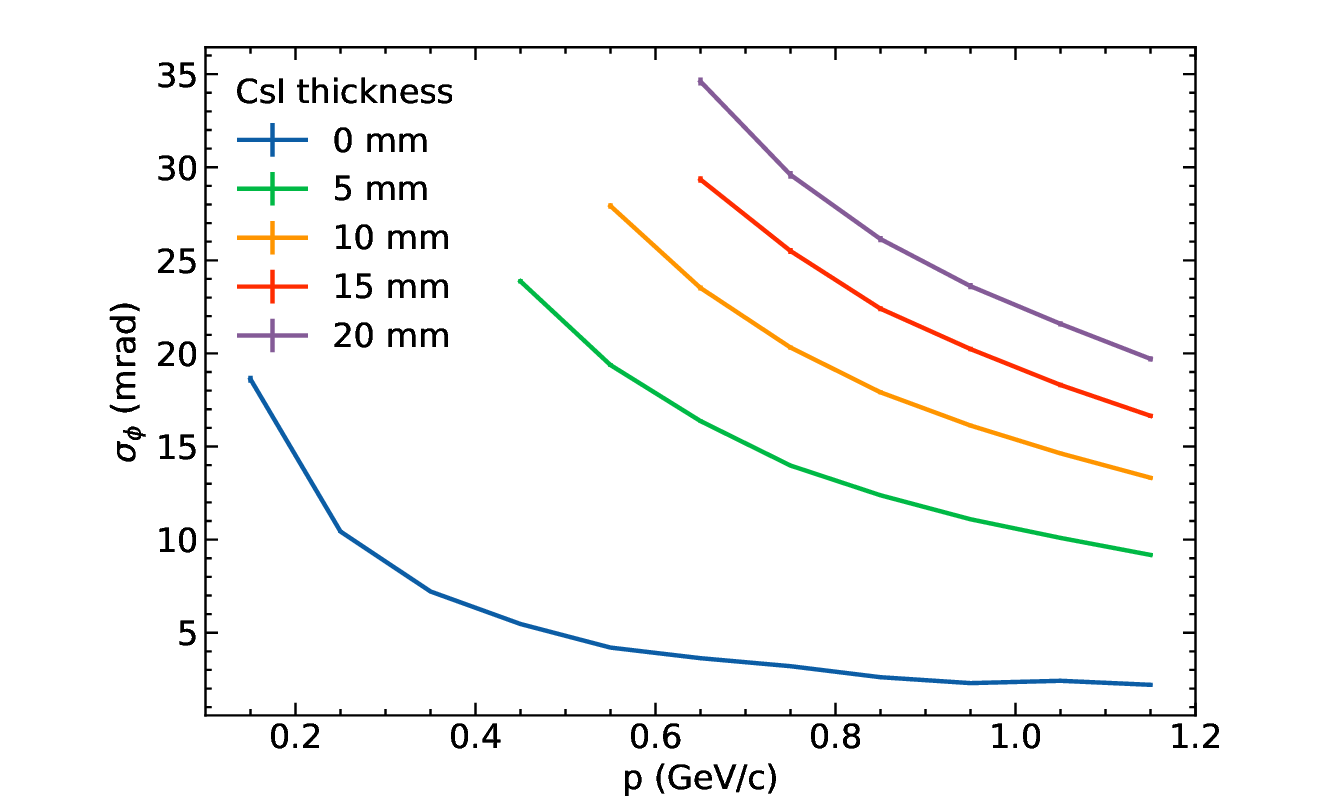}}
        	\caption{Azimuth angle resolution, \phires, as a function of of momentum p  for (a) protons (b) pions generated within the range of $0.1 < \cos\theta < 0.2$ for adding a CsI layer with the thickness of 0, 5, 10, 15, and 20 mm. }
        	\label{fig:phi_resolution}
        \end{figure}

The momentum resolution~(\pres), the polar angle resolution~(\tres), and the azimuthal angle resolution~(\phires) are also checked with respect to the track momentum 
for different thickness of the added CsI tube. Figure~\ref{fig:p_resolution} shows \pres\ of \pp\ and \pim. We can see that additional CsI mainly affects the \pp\ with momentum lower than 0.8~GeV/$c$, but for \pim, 20~mm thick CsI worsens the \pres\ by less than 1~MeV/$c$ in the full momentum range. The CsI effect on \tres\ and \phires\ for \pp\ and \pim\ is shown in Fig.~\ref{fig:t_resolution} and Fig.~\ref{fig:phi_resolution}. Without additional CsI, the \tres\ and \phires\ is below 10~milliradians for the most of momentum range.  Additional CsI material has a greater impact for the low momentum tracks. With the increase of CsI thickness, the impact of adding the same additional thickness of CsI decreases gradually. 
With the thickness of additional CsI increasing, the tracking efficiency of low momentum particles is low, thus it is more difficult to obtain accurate \pres, \tres\ and \phires\ (missing points in  Figs.~\ref{fig:p_resolution},~\ref{fig:t_resolution} and \ref{fig:phi_resolution}).
        
\subsection{Study of single antineutron sample}
        
Although not very reliable, we may still get a rough feeling about the probability of the antineutron and CsI interactions by simulating a single antineutron and adding CsI tube with different thickness in the detector. The range of antineutron momentum is set to $[0,~1.2]$~GeV/$c$ while the polar angle is the same as that of \pp\ and \pim. In total, $10^6$ events of \nbar sample are simulated. We can get the fractions of \nbar\ annihilation and \nbar-nuclei elastic scattering in the additional CsI. For \nbar\ annihilation event, the final position of the antineutron indicates where the annihilation happened, it can be obtained from the MC truth information; for \nbar-nuclei elastic scattering event, the momentum difference between before entering and after leaving the additional CsI indicates whether or not elastic scattering has occurred. The threshold of momentum change is set to 0.01~GeV/$c$ although the momentum change must be strictly equal to zero if there is no scattering happened.
        
The above method of identifying the annihilation and elastic scattering events is validated with the simulation of antineutron and beam pipe interactions. The simulation shows that there are 4.7\% events, including 3.7\% of annihilation and 1.0\% of elastic scattering, of antineutron and beam pipe interactions happened. If the threshold of momentum is increased to 0.3~GeV/$c$, the results become 2.6\%, 1.8\%, and 0.8\%, respectively, which are in consistency with the estimation of 1--2\% in Ref.~\cite{hypronProjectileFromJpsi}. 

\begin{figure}[htbp]
        	\centering  
        	\subfigbottomskip=2pt 
        	\subfigcapskip=-5pt 
        	\subfigure[]{
        		\includegraphics[width=0.45\linewidth]{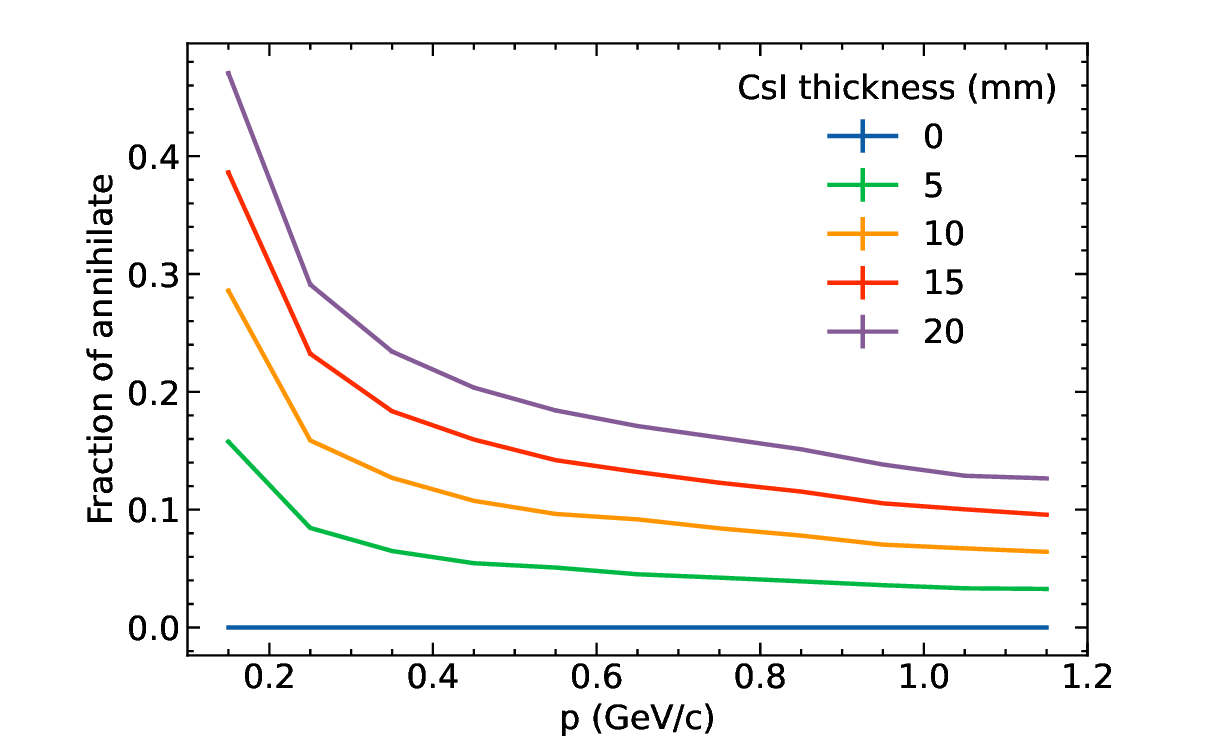}}
        	\subfigure[]{
        		\includegraphics[width=0.45\linewidth]{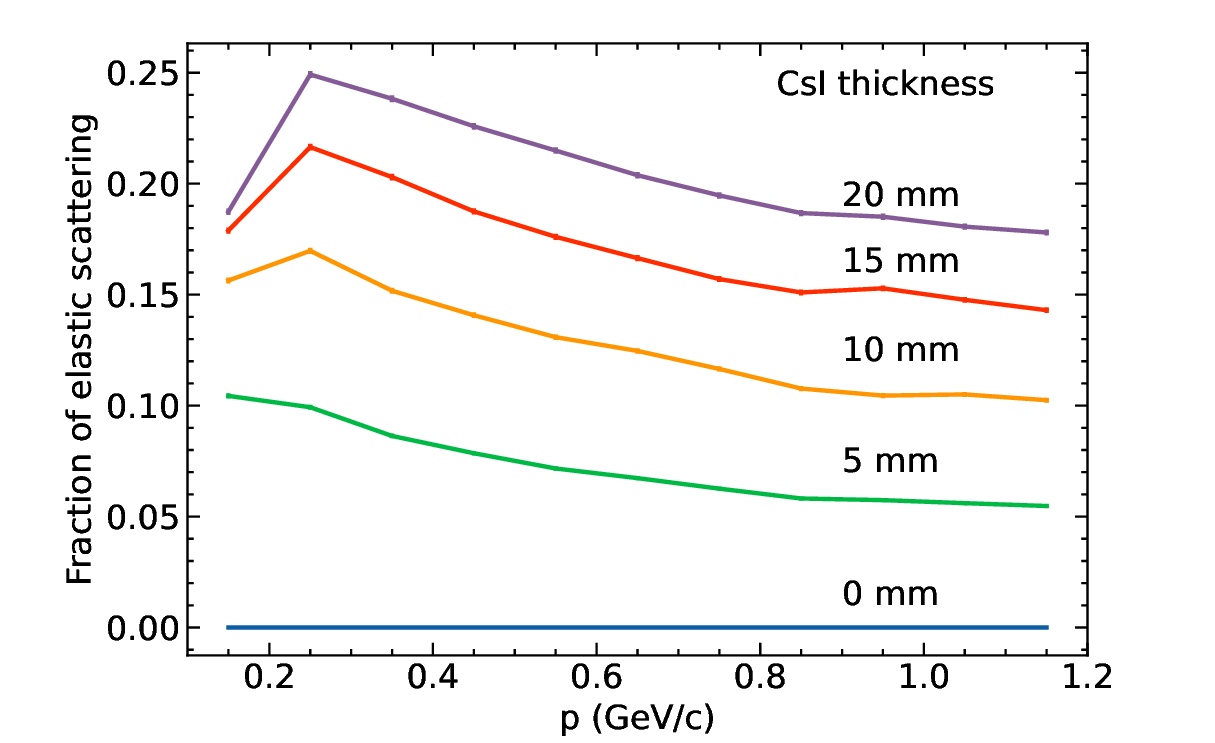}}
        	\caption{Probability of interactions (a) annihilation (b) elastic scattering between anti-neutron and CsI as a function of momentum for adding a CsI layer with the thickness of 0, 5, 10, 15, and 20 mm. }
        	\label{fig:fraction_nbar}
        \end{figure}

Figure~\ref{fig:fraction_nbar} shows fractions of \nbar\ annihilation and \nbar-nuclei elastic scattering in the additional CsI tube with thickness of 0, 5, 10, 15, and 20~mm. We find that the fractions increase almost linearly as a function of the thickness of the additional CsI tube, and adding every 1~mm of the CsI results in 2.5\% of antineutron interaction, of which 1.2\% is annihilation and 1.3\% is elastic scattering. The ratio of elastic scattering and annihilation is different from that in the beam pipe. 
        
With a 10~mm thick CsI tube added, we find that about 24\% of the antineutrons interact with it, of which 11\% are annihilation and 13\% are elastic scattering. The chance for multiple interactions will be increased if more CsI material is added and this may make the study of the antineutron interaction more complicated. This should be considered in the design of the experiment. 

\subsection{\boldmath Study of the $J/\psi\to $\pp\pim\nbar~sample}

MC samples of $J/\psi\to $\pp\pim\nbar\ are simulated with additional CsI tube of thickness 0, 5, 10, 15, and 20~mm and subject to event selection to tag the antineutrons. Two charged tracks with opposite charge and originating from the interaction point are required in each event, after particle identification, they are marked as \pp\ and \pim. The recoil mass of \pp\ and \pim\ is required to be between 0.90 and 0.98~GeV/$c^2$. In total, $10^6$ events of $J/\psi\to $\pp\pim\nbar\ sample are simulated. 

\begin{figure}[htbp]
	\centering  
	\subfigbottomskip=2pt 
	\subfigcapskip=-5pt 
	\subfigure[]{
	    \includegraphics[width=0.45\linewidth]{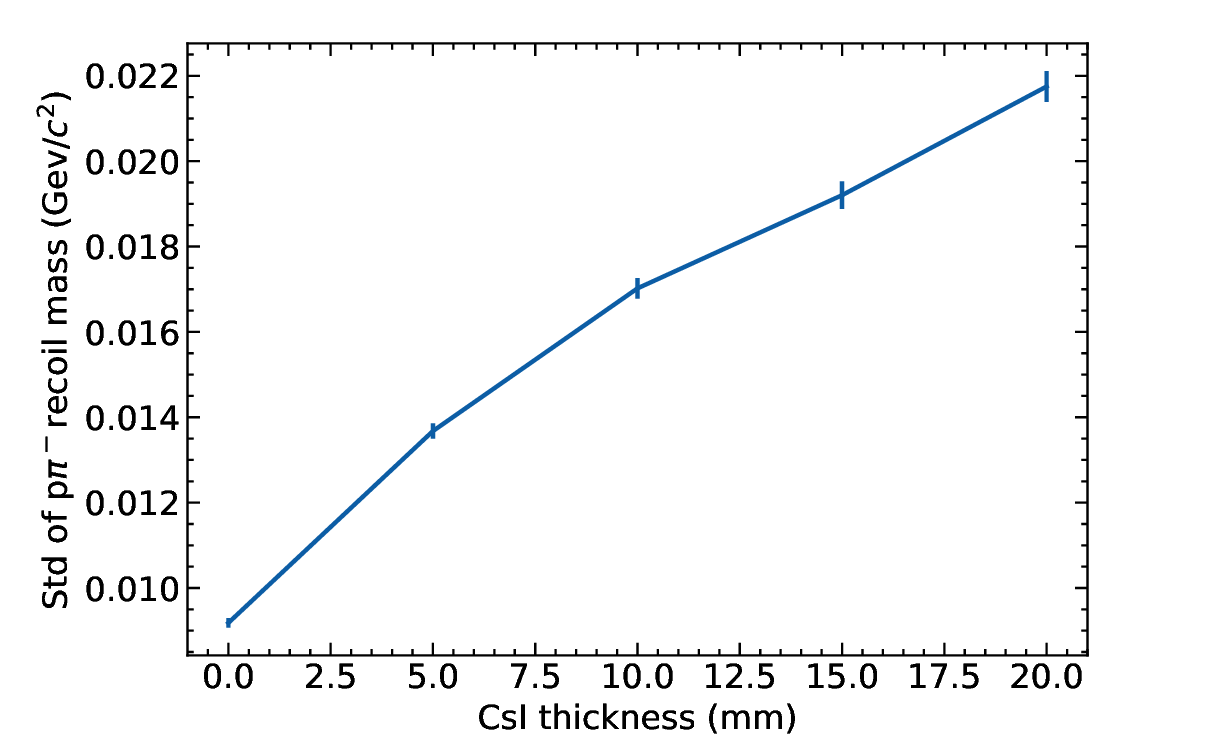}}
	\subfigure[]{
		\includegraphics[width=0.45\linewidth]{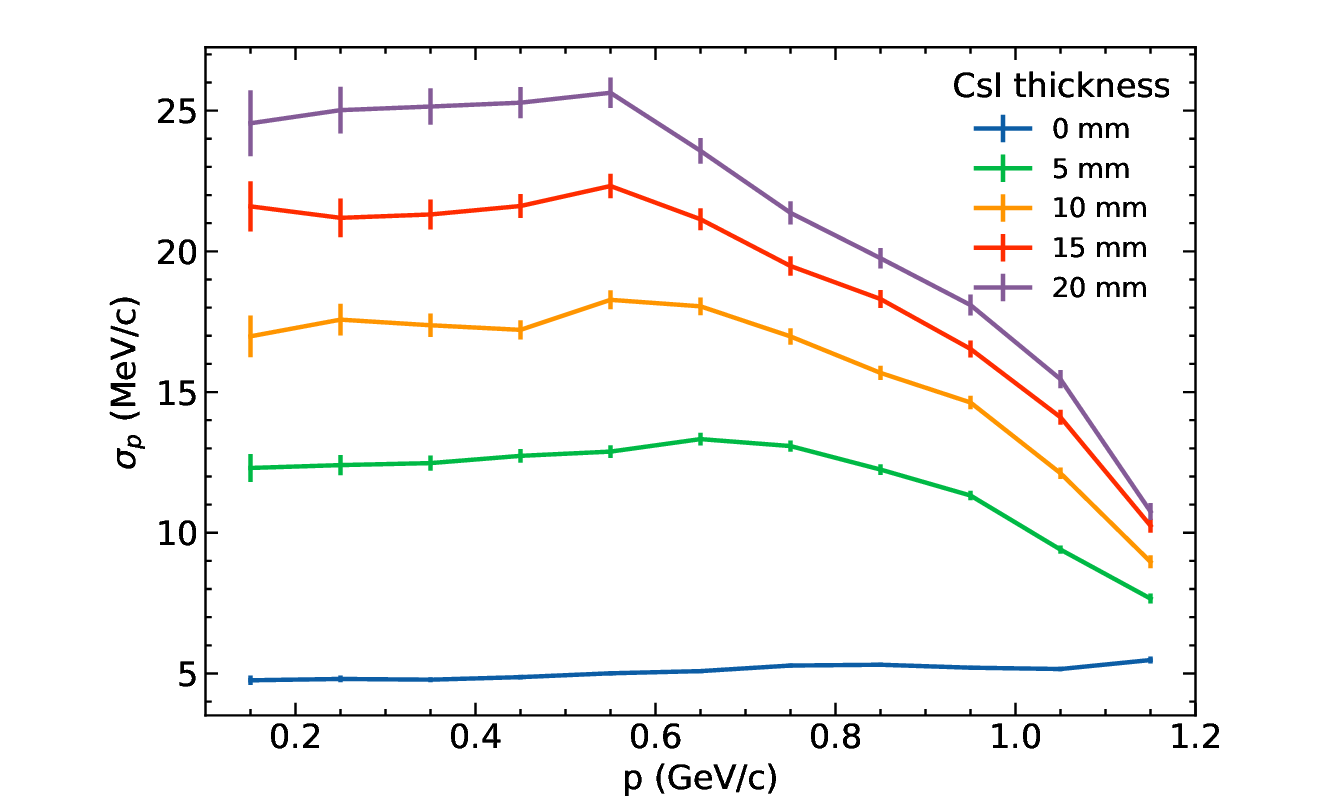}}
	\subfigure[]{
		\includegraphics[width=0.45\linewidth]{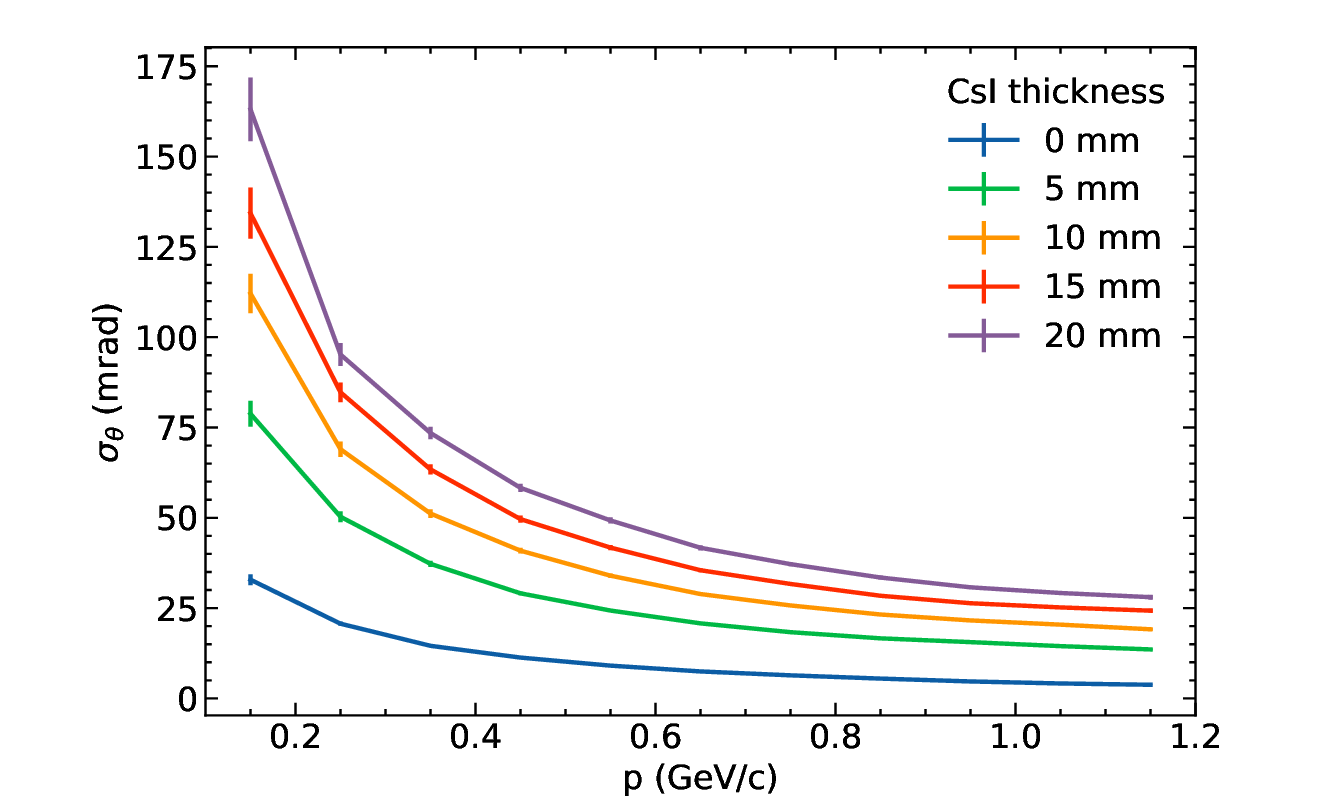}}
	\subfigure[]{
		\includegraphics[width=0.45\linewidth]{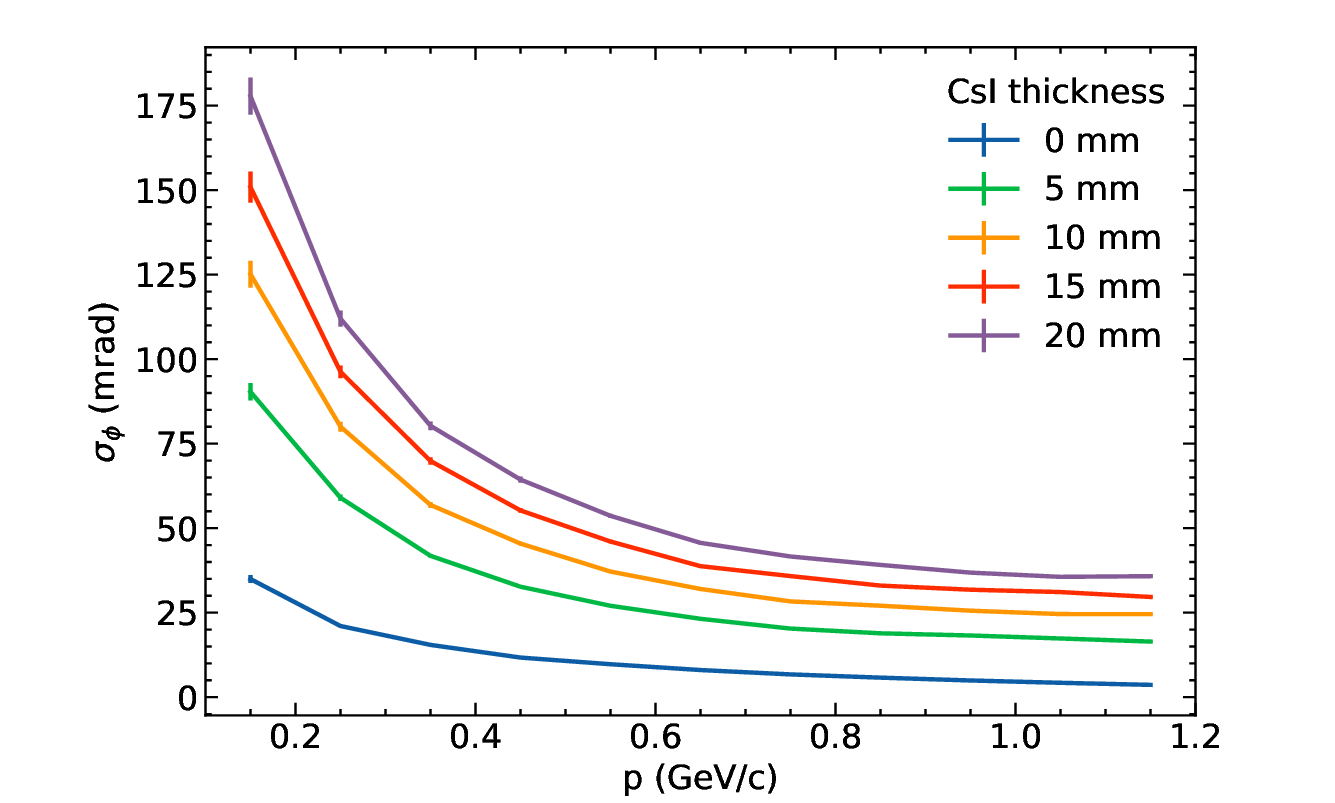}}
	\caption{Differences between the reconstructed and true values (a) shows the resolution of recoil mass of \pp and \pim as a function of the thickness of additional CsI layer. (b) (c) (d) show respectively \pres, $\sigma_\theta$ and $\sigma_\phi$ (d) of the recoiling antineutron as a function of momentum for adding a CsI layer with the thickness of 0, 5, 10, 15, and 20 ~mm.}
	\label{fig:nbar_resolutions}
\end{figure}

The additional CsI has an obvious influence on the resolutions of the recoiling antineutrons. After thicker CsI tube is added, the distribution of the recoiling mass is significantly widened, thus the resolution of the recoiling mass is reduced as shown in Fig.~\ref{fig:nbar_resolutions}(a). Figure~\ref{fig:nbar_resolutions}(b, c, d) shows \pres, \tres\ and \phires\ of the tagged antineutrons without or with different thickness additional CsI tube added, they decrease with the increase of momentum.

\begin{figure}[htbp]
	\centering  
	\subfigbottomskip=2pt 
	\subfigcapskip=-5pt 
	\subfigure[]{
		\includegraphics[width=0.45\linewidth]{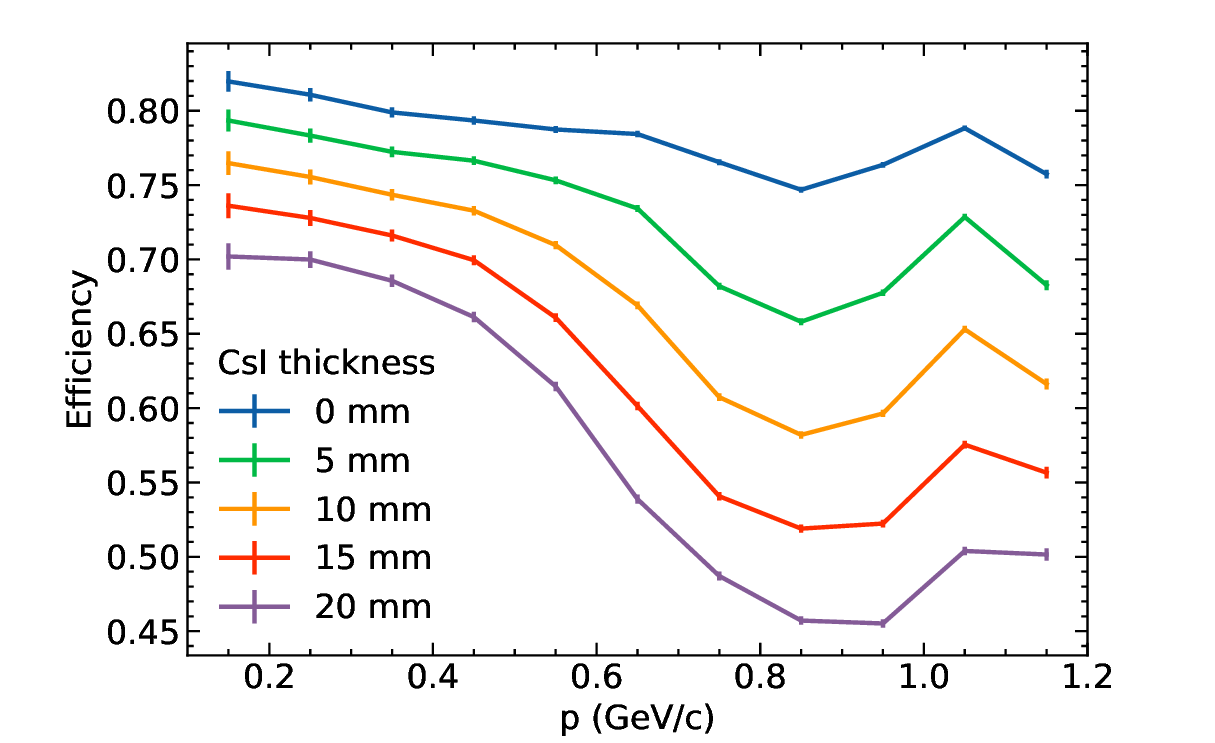}}
	\subfigure[]{
		\includegraphics[width=0.45\linewidth]{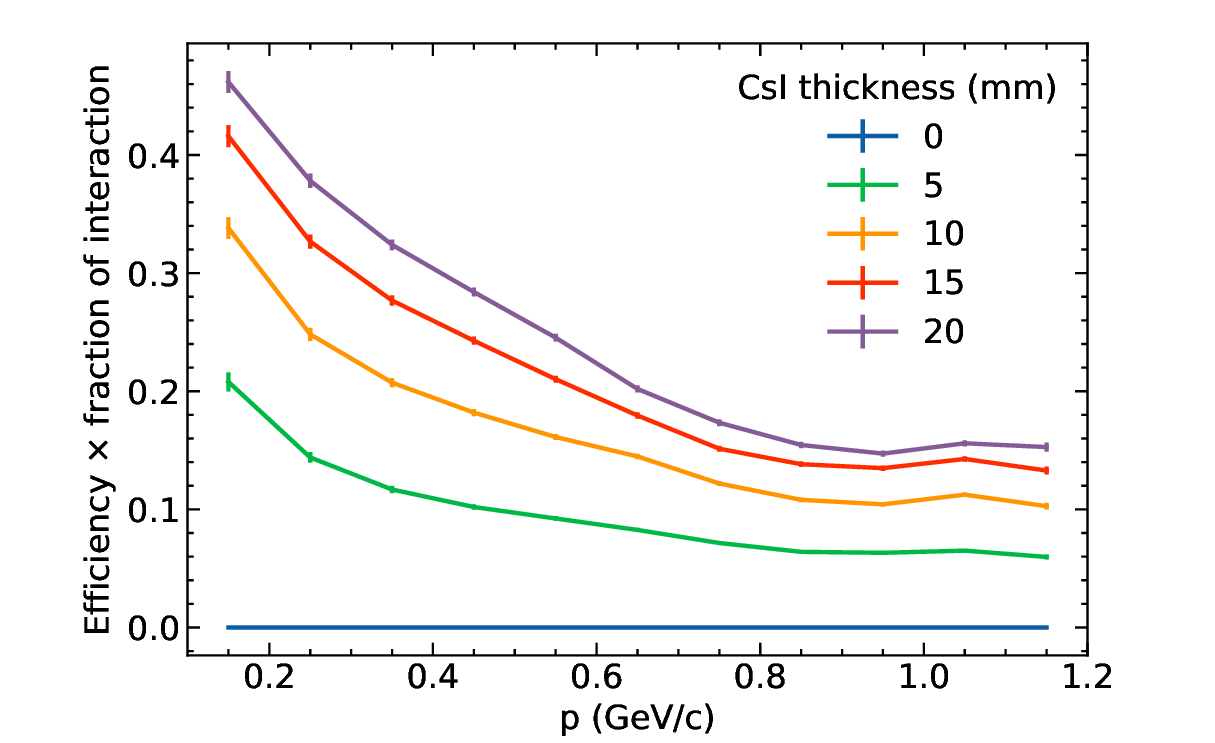}}
	\subfigure[]{
		\includegraphics[width=0.45\linewidth]{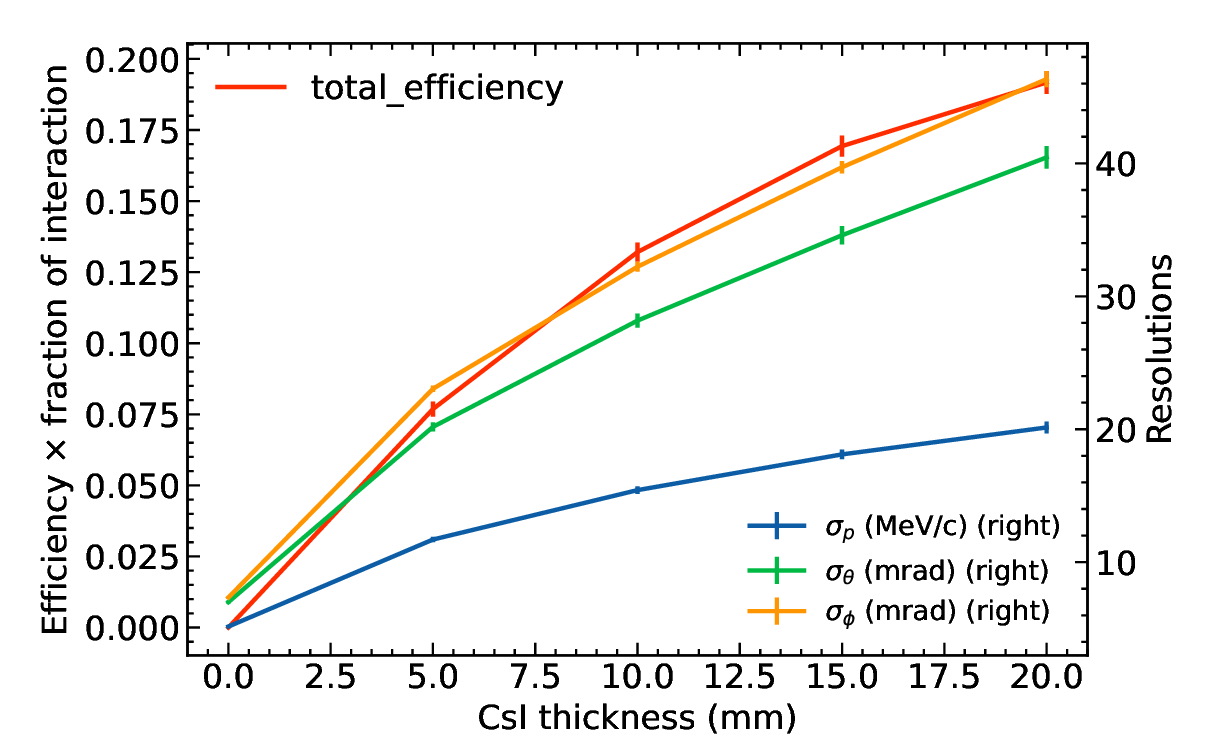}}
		\subfigure[]{
		\includegraphics[width=0.45\linewidth]{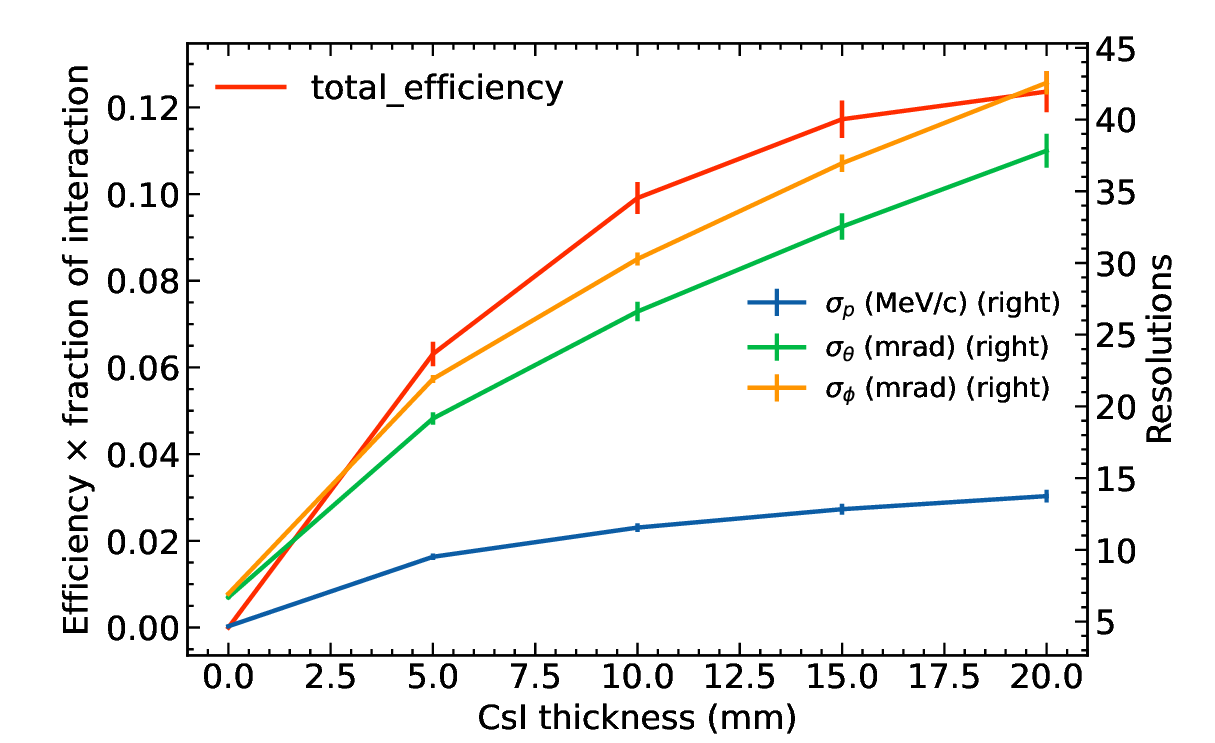}}
	\caption{(a) (b) respectively show antineutron selection efficiencies and yields in different momentum ranges for adding a CsI layer with the thickness of 0, 5, 10, 15, and 20 ~mm. (c) The variation trends for antineutron yield, \pres\, \tres\ and \phires\ (d) as a function of the thickness of the additional CsI layer. For the left y-axis, the unit is percent. For the right y-axis, the unit for \pres\ is MeV/$c$ and the units for \tres\ and \phires\ are mrad; (d) shows the similar results to (c) after applying stricter select criteria. }
	\label{fig:ppinbar}
\end{figure}

With additional CsI of different thickness, the selection efficiencies of the recoiling antineutron in different momentum ranges are shown in Fig.~\ref{fig:ppinbar}(a). The efficiency decreases almost uniformly as the CsI thickness increases and the low valley in the range of 0.8 to 1.0~GeV/$c$ is due to that the corresponding momentum of the tagging proton can be very low and can hardly be detected.
When a 10~mm thick CsI tube added, the total efficiency is 67.5\%, which is reduced by 10.8\% compared to 78.3\% without additional CsI; when a 20~mm thick CsI tube is added, the total efficiency and the reduction become 57.3\% and 21.0\%.
        
With the recoiling antineutron efficiency and the fraction of interactions between antineutron and additional CsI, the fraction of the antineutrons that can be observed is obtained by simply multiplying them. Although the efficiency decreases, the antineutron interactions that can be selected still increases monotonously, as shown in Fig.~\ref{fig:ppinbar}(b). However, the growth rate slows down as the CsI thickness increases.
        
Figure~\ref{fig:ppinbar}(c) shows the total efficiency of antineutron and CsI interactions and the resolutions vary with the thickness of the CsI tube. We also studied the changes in total efficiency and resolution when using stricter selection conditions. We have added the following conditions: the distance between the vertex and the xy-plane is less than 0.5cm, while the distance from the z-axis is less than 5cm, the fitted chi2 of the vertex is less than 30, and the recoil mass is between 0.92 and 0.96~Gev/c. It can be seen that when the thickness of CsI is greater than 15~mm, the total efficiency almost does not increase, while the resolution still decreases linearly.

\section{Conclusion and discussion}
\label{sec:conclusion}

From the MC simulation of single proton, \pim\, and \nbar, as well as $J/\psi\to p \pi^- \bar{n}$ samples, without or with additional CsI tube of different thickness, we find that adding thicker CsI tube can produce and select more antineutron-CsI interactions. However, the \pres, \tres\ and \phires\ of the antineutrons will decrease and make the measurement of the interactions less precise. So the optimization of the thickness of the additional CsI tube is a compromise between event rate and the resolutions of the antineutron source (Fig.~\ref{fig:ppinbar}(c)). Therefore, we can draw the following conclusion. Firstly, based on the specific research content of the experiment, a reasonable thickness range can be obtained by referring to the Figure (Fig.~\ref{fig:ppinbar}). In order to ensure the accuracy of the antineutron tagging, the thickness of CsI should not more than 15~mm. Secondly, the efficiency and resolution of tag antineutrons vary significantly in different momentum ranges. When studying antineutrons with high momentum (above 0.8~GeV/c), thicker CsI needs to be added to ensure the quantity of tagged antineutrons. Although thicker CsI will results in more interactions recorded, the events may suffer from multiple interactions thus make the study more complicated. The reconstruction and identification of the \nbar\ and CsI interaction events should be investigated further and it is out of the scope of this work. Meanwhile, when selecting the thickness of CsI, the duration of the experiment should be taken into account. Prolonged time can result in some uncertainty, as the state of the detector changes over time.

In BESIII experiment, with the ability of generating 10 billion \jpsi\ events per year, we estimate that when the thickness of the additional CsI tube is 10~mm, 2.6 million events of interaction between antineutron and CsI tube can be collected per year according to the total efficiency (Fig.~\ref{fig:ppinbar}(d)). The most important thing is that the sample contains a large number of samples above 500~MeV, which is unique in the world~\cite{hypronProjectileFromJpsi}. This unprecedented data sample will benefit the antineutron simulation very significantly.
    
    
There is room for improvement in the experimental design. With the increase of $\cos\theta$, the efficiency and resolution of \pp, $\pi^-$, and antineutron decrease significantly, because the particles need to pass through thicker CsI material before entering the detector. To ensure that the particle at different polar angle passes through the same amount of material, the thickness of additional CsI tube can be different at different polar angle, for example, to be $t_0\sin\theta$, where $t_0$ is the thickness at $\theta=90^\circ$.
    
The study presented in this article is easy to be extended to other materials, especially for those solid materials like other kinds of crystals, such as NaI, PbWO$_4$, and BGO, used as electromagnetic calorimeter material, or iron and lead used as hadronic calorimeter material in high energy and nuclear physics experiments. 

    
A similar method can be used to study the interaction between hyperons and nucleons. Many hyperons can be produced and tagged from \jpsi\ decays as proposed in Ref.~\cite{hypronProjectileFromJpsi}. In this case, the target must be changed into liquid hydrogen or liquid deuterium. We also simulated the efficiency and resolutions of \pp\ and \pim\ by adding a layer of liquid hydrogen and liquid deuterium contained between two solid tubes. Limited by the space inside the BESIII detector~\cite{bes3}, only about 20~mm of matter can be added at most, which is far from enough for the study of hyperon and nucleon interactions. Future experiments plan to do such kinds of measurements may design beam pipe with smaller radius (10~mm or less rather than 33.7~mm at BESIII) to allow more space for the light target material. 
    
\section{Acknowledgments}

Thank Weimin Song and Liang Liu for their help with the software.
This work is supported in part by National Key Research and Development Program of China under Contract No.~2020YFA0406300, National Natural Science Foundation of China (NSFC) under contract No.~12275297.


\begin{thebibliography}{40}

\bibitem{pdg}
R.~L.~Workman \textit{et al.} (Particle Data Group),
PTEP \textbf{2022}, 083C01 (2022)
doi:10.1093/ptep/ptac097

\bibitem{CLEO:2008aum}
S.~B.~Athar \textit{et al.} (CLEO Collaboration),
Phys. Rev. Lett. \textbf{100}, 181802 (2008).

\bibitem{BESIII:2012imn}
M.~Ablikim \textit{et al.} (BESIII Collaboration),
Phys. Rev. D \textbf{86}, 052011 (2012).

\bibitem{BESIII:2021tbq}
M.~Ablikim \textit{et al.} (BESIII Collaboration),
Nature Phys. \textbf{17}, no.11, 1200-1204 (2021).

\bibitem{bes3} M.~Ablikim {\it et al.} (BESIII Collaboration), Nucl.\ Instrum.\ Methods Phys. Res., Rect.\ A {\bf 614}, 245 (2010).

\bibitem{cleoc}
R.~A.~Briere \textit{et al.} (CLEO Collaboration),
CLNS-01-1742.

\bibitem{belle2}
E.~Kou \textit{et al.} (Belle-II Collaboration),
PTEP \textbf{2019}, no.12, 123C01 (2019)
[erratum: PTEP \textbf{2020}, no.2, 029201 (2020)].

\bibitem{geant4} S.~Agostinelli {\it et al.}, (GEANT4 Collaboration), Nucl.\ Instrum.\ Meth.\ A {\bf 506}, 250 (2003). 

\bibitem{Liu:2021rrx}
L.~Liu, X.~Zhou and H.~Peng,
Nucl. Instrum. Meth. A \textbf{1033}, 166672 (2022).

\bibitem{E767} 
T.~Armstrong \textit{et al.} [BROOKHAVEN-HOUSTON-PENNSYLVANIA STATE-RICE],
Phys. Rev. D \textbf{36}, 659-673 (1987).

\bibitem{OBELIX} 
M.~Agnello 
\textit{et al.},
Nucl. Instrum. Meth. A \textbf{399}, 11-26 (1997).

\bibitem{nbarPhysics} 
T.~Bressani and A.~Filippi,
Phys. Rept. \textbf{383}, 213-297 (2003).

\bibitem{hypronProjectileFromJpsi} 
C.~Z.~Yuan and M.~Karliner,
Phys. Rev. Lett. \textbf{127}, no.1, 012003 (2021).

\bibitem{bepc2} Q. Qin, L. Ma, J. Wang, and C. Zhang,  Conf. Proc. C {\bf 100523}, 2359 (2010), IPAC-2010-WEXMH01, http://accelconf.web.cern.ch/AccelConf/IPAC10/papers/wexmh01.pdf.

\bibitem{bes3physics} 
D.~M.~Asner {\it et al.}, Int.\ J.\ Mod.\ Phys.\ A {\bf 24}, S1 (2009).

\bibitem{bes3physicsFuture} 
M.~Ablikim \textit{et al.} (BESIII Collaboration),
Chin. Phys. C \textbf{44}, no.4, 040001 (2020).

\bibitem{boss}
 W.~Li {\it et al.}, Proc. Int. Conf. Comput. High Energy and Nucl. Phys. 225 (2006).

\bibitem{bes3hough}
J.~Zhang  {\it et al.}, Radiat. Detect. Technol. Methods {\bf 2}, 20 (2018).
 
\bibitem{kalFit}
 J. Wang {\it et al.}, Chin. Phys. C {\bf 33}, 870-879 (2009)

\end{thebibliography}
\end{document}